\documentclass[aip,pof,showpacs,superscriptaddress,groupedaddress]{revtex4-1} 

\usepackage{graphicx}        
\usepackage{dcolumn}        
\usepackage{bm}             
\usepackage{amssymb}
\usepackage{amsmath}        
\usepackage{color}
\definecolor{korr_26Apr}{rgb}{0,0,0} 
\definecolor{red}{rgb}{1,0,0}

\def \d{\mathrm{d}}
\def \D{\mathrm{D}}

\begin{document}

\widetext

\title{The fluctuation energy balance in non-suspended fluid-mediated particle transport}
\author{Thomas P\"ahtz$^{1,2}$}
\email{0012136@zju.edu.cn}
\author{Orencio Dur\'an$^3$, Tuan-Duc Ho$^4$, Alexandre Valance$^4$, Jasper F. Kok$^5$}
\affiliation{1.~Institute of Physical Oceanography, Ocean College, Zhejiang University, 310058 Hangzhou, China \\
2.~State Key Laboratory of Satellite Ocean Environment Dynamics, Second Institute of Oceanography, 310012 Hangzhou, China \\
3.~MARUM-Center for Marine Environmental Sciences, University of Bremen, 28359 Bremen, Germany \\
4.~Institut de Physique de Rennes, UMR UR1-CNRS 6251, Universit\'e de Rennes 1, 35042 Rennes Cedex, France \\
5.~Department of Atmospheric and Oceanic Sciences, University of California, Los Angeles, 90095-1565 Los Angeles, CA, USA}

\begin{abstract}

Here we compare two extreme regimes of non-suspended fluid-mediated particle transport, transport in light and heavy fluids (``saltation'' and ``bedload'', respectively), regarding their particle fluctuation energy balance. From direct numerical simulations, we surprisingly find that the ratio between collisional and fluid drag dissipation of fluctuation energy is significantly larger in saltation than in bedload, even though the contribution of interparticle collisions to transport of momentum and energy is much smaller in saltation due to the low concentration of particles in the transport layer. We conclude that the much higher frequency of high-energy particle-bed impacts (``splash'') in saltation is the cause for this counter-intuitive behavior. Moreover, from a comparison of these simulations to Particle Tracking Velocimetry measurements which we performed in a wind tunnel under steady transport of fine and coarse sand, we find that turbulent fluctuations of the flow produce particle fluctuation energy at an unexpectedly high rate in saltation even under conditions for which the effects of turbulence are usually believed to be small.

\end{abstract}

\maketitle

\section{Introduction}
Fluid-mediated particle transport occurs when a fluid flows sufficiently strongly over a particle bed. The two most prominent examples, water flow over a river bed and air flow over a sand desert, constitute two extreme regimes of fluid-mediated particle transport. These regimes are characterized by values of the particle-fluid density ratio, $s=\rho_{p}/\rho_{f}$, which are either sufficiently close to unity (e.g., $s\approx2.65$ for sand in water) or sufficiently far from unity (e.g., $s\approx2250$ for sand in air) \cite{Duranetal12}, where $\rho_{p}$ and $\rho_{f}$ are the particle and fluid density, respectively. If particle transport in suspension with the fluid is much weaker than particle transport along the surface, which is typical for sand transport in air \cite{Bagnold41} and sand and gravel transport in mildly-sloped rivers \cite{vanRijn93}, these two regimes are known as ``bedload'' and ``saltation'', respectively \cite{Duranetal12}.

Bedload and saltation are responsible for a wide variety of geophysical phenomena, including wind erosion, dust aerosol emission, and the formation of dunes and ripples on ocean floors, river beds, and planetary surfaces \cite{Bagnold41,vanRijn93,Garcia07,Shao08,Bourkeetal10,Duranetal11,Koketal12}. While many aspects of bedload and saltation have been investigated in numerous experimental and theoretical studies in the last century (e.g., \cite{Bagnold41,Einstein50,Kawamura51,Bagnold56,Yalin63,Owen64,Bagnold66,AshidaMichiue72,Bagnold73,EngelundFredsoe76,FernandezLuqueBeek76,Kind76,LettauLettau78,UngarHaff87,Sorensen91,NinoGarcia98a,Cheng02,Andreotti04,Sorensen04,AbrahamsGao06,DuranHerrmann06,Creysselsetal09,KokRenno09,Lajeunesseetal10,Carneiroetal11,Duranetal11,Hoetal11,Duranetal12,Laemmeletal12,Paehtzetal12,Carneiroetal13,Paehtzetal13,Duranetal14a,Duranetal14b,Hoetal14,Paehtzetal14,Schmeeckle14}), a theoretical concept unifying bedload and saltation, which includes predictions for intermediate regimes (e.g., $s\approx100$) \cite{Duranetal12}, is still missing.

In order to achieve such a unifying understanding, it is essential to first characterize the fundamental physical differences between both regimes. Here we discovered that one such fundamental difference is the manner in which particle fluctuation energy, also known as ``granular temperature'', is dissipated. In fact, in steady particle shear flows, such as fluid-mediated particle transport, the continuous conversion of mean energy into fluctuation energy through granular shear work \cite{Campbell06} must be balanced by dissipation through interparticle collisions as well as fluid drag, which dissipates fluctuation energy because it tends to align the velocities of all particles by changing them towards the velocity of the ambient fluid. Fluid drag dissipation in turn can be separated into dissipation through mean drag and production due to turbulent fluctuations of the flow speed.

Here we investigate the particle fluctuation energy balance through Discrete Element Method (DEM) simulations of steady, homogeneous non-suspended fluid-mediated particle transport using the model described in Ref.~\cite{Duranetal12}. Indeed, as shown in very detail in the Supplementary Material \cite{Suppl_Mat} as well as in Refs.~\cite{Babic97,Goldhirsch10}, it is possible to derive the average particle fluctuation energy balance only from Newton's axioms and to determine all terms appearing in this balance from the simulation data. From this investigation, we find that a significantly larger fraction of fluctuation energy is dissipated through interparticle collisions in saltation than in bedload. In particular, in the region near the particle bed, collisional dissipation dominates fluid drag dissipation in saltation, in contrast to bedload where fluid drag dissipation prevails. As a side effect of this and due to their tendency to dissipate granular temperature anisotropies (defined below), interparticle collisions in saltation convert horizontal fluctuating energy into vertical one at a rate which is higher than the rate of dissipation through fluid drag in the vertical direction. These finding are surprising, considering the predominance of interparticle collisions within the dense transport layer in bedload, if compared with the dilute transport layer in saltation \cite{Duranetal12}.

Moreover, a comparison of these DEM simulations to Particle Tracking Velocimetry (PTV) measurements, which we performed in a wind tunnel under steady transport of fine (mean particle diameter, $d\approx230\mu$m) and coarse sand ($d\approx630\mu$m), indicates that turbulent fluctuations of the flow produce fluctuation energy at an unexpectedly high rate in saltation. In particular, turbulent production of vertical fluctuation energy seems to surpass both the contributions from interparticle collisions and mean drag by a large margin. This is surprising since the effects of turbulence on saltation of sand are usually believed to be small, especially for coarse sand \cite{KokRenno09}.

\section{Discrete Element Model simulations}
We simulated very simple, but relevant conditions: steady, homogeneous particle transport. That is the transport of particles along a horizontal, flat, infinite, homogeneous particle bed subjected to a unidirectional, steady, fluid flow. Thereby each particle was modeled by a sphere with a constant density $\rho_{p}$ and a diameter $d_{p}$ uniformly distributed between $0.8$ and $1.2$ times the mean diameter, $d$, while the flow was modeled by a Newtonian fluid with a density $\rho_{f}$, a kinematic viscosity $\nu$, and a constant fluid shear stress $\tau$ (i.e., inner turbulent boundary layer flow \cite{Duranetal12}). The mean velocity of the turbulent flow was modeled using a modified version of the mixing length theory, in which the turbulent mixing length is modified by the
presence of the particle phase \cite{Duranetal12}, while turbulent fluctuations of the mean flow were neglected. Moreover, interparticle collisions were taken into account using contact dynamics \cite{Duranetal12}, and we further considered the two most important fluid-particle interactions: buoyancy and fluid drag. The entire system was under gravity with a gravitational constant $g$ and the simulations where two-dimensional (note that 2D and 3D simulations yield qualitatively similar results \cite{Carneiroetal11,Carneiroetal13}). We refer to Dur\'an et al. \cite{Duranetal12} for further modeling and simulation details.

Steady, homogeneous particle transport can be characterized by three dimensionless numbers \cite{Duranetal12}: the Shields number ($\Theta$), the particle Reynolds number ($\mathrm{Re}$), and the density ratio ($s$),
\begin{eqnarray}
 \Theta&=&\frac{\tau}{(\rho_{p}-\rho_{f})gd}, \\
 \mathrm{Re}&=&\frac{d}{\nu}\sqrt{(s-1)gd}, \label{defRe} \\
 s&=&\frac{\rho_{p}}{\rho_{f}}.
\end{eqnarray}
Simulations were performed with fixed $\mathrm{Re}=10$ and for two values of $s$, $s=2$ (bedload) and $s=2000$ (saltation). We note that this particle Reynolds number would correspond to $d\approx183\mu$m for transport in water ($s=2.65$, $\nu=0.001\mathrm{m^2/s}$) and to $d\approx98\mu$m for transport in air ($s=2208$, $\nu=1.43\times10^{-5}\mathrm{m^2/s}$). Moreover, $\Theta$ was varied between one and several times the transport threshold Shields number, $\Theta_t$, which is defined as the minimal Shields number at which transport once initiated can be sustained. For $s=2$ and $\mathrm{Re}=10$, we obtained $\Theta_t=0.12$ and for $s=2000$ and $\mathrm{Re}=10$, $\Theta_t=0.004$ from the simulations. The simulations were run sufficiently long to ensure that the system was in the steady state for the major part of time. We note that velocity-based quantities, such as $\mathrm{Re}$ in Eq.~(\ref{defRe}), are henceforth based on the velocity scale $\sqrt{(s-1)gd}$ \cite{Duranetal12} because this definition ensures that the average particle velocity in units of $\sqrt{(s-1)gd}$ under threshold conditions ($\Theta=\Theta_t$) has roughly the same value irrespectively of the value of $s$.

\section{Fluctuation energy balance}
Our simulations allowed us to investigate the balance of the granular temperature ($T$), which is formally defined by $T=\frac{1}{N_{d}}T_{ii}$ (Einsteinian summation) and $T_{ij}=\langle c_ic_j\rangle$, where $N_{d}$ is the number of dimensions ($2$ in our simulations), $\mathbf{v}$ the particle velocity, $\langle\cdot\rangle$ a local mass-weighted ensemble average, and $\mathbf{c}=\mathbf{v}-\langle \mathbf{v}\rangle$ the fluctuation velocity. Moreover, by defining a Cartesian coordinate system $(x,z)$ with $x$ in direction parallel and $z$ in direction normal to the particle bed, $T_{xx}$ and $T_{zz}$ become the aforementioned horizontal and vertical granular temperatures, respectively. For steady, homogeneous particle transport ($\partial/\partial t=\partial/\partial x=0$), the fluctuation energy balances read (for the derivation, see Supplementary Material \cite{Suppl_Mat})
\begin{eqnarray}
 0=\frac{1}{2}\rho\frac{\D T_{xx}}{\D t}&=&-\frac{\d q_{zxx}}{\d z}+W^\mathrm{shear}-\Gamma^\mathrm{drag}_{xx}-\Gamma^\mathrm{coll}_{xx}, \label{energyx} \\
 0=\frac{1}{2}\rho\frac{\D T_{zz}}{\D t}&=&-\frac{\d q_{zzz}}{\d z}-\Gamma^\mathrm{drag}_{zz}-\Gamma^\mathrm{coll}_{zz}, \label{energyz}
\end{eqnarray}
where $\rho$ is local particle mass per unit volume, $\D/\D t=\langle v_z\rangle\d/\d z$ is the material derivative (i.e., the time derivative in the reference frame moving with the particle flow), which vanishes since the flux of upward-moving particles must exactly compensate the flux of downward-moving particles in steady, homogeneous particle transport ($\langle v_z\rangle=0$ \cite{Suppl_Mat}). Moreover, the fluctuation energy flux tensor ($q_{ijk}$), hereafter called ``granular heat flux'' tensor, the granular shear work ($W^\mathrm{shear}$), and the fluctuation energy dissipation rate tensors due to fluid drag ($\Gamma^\mathrm{drag}_{ij}$) and interparticle collisions ($\Gamma^\mathrm{coll}_{ij}$) are given by \cite{Suppl_Mat}
\begin{eqnarray}
 q_{ijk}&=&q_{ijk}^\mathrm{c}+q_{ijk}^\mathrm{t}, \label{def_q} \\
 q_{ijk}^\mathrm{c}&=&\frac{1}{2}\overline{\sum_{mn}F_{j}^{mn}c_{k}^m(x_i^m-x_i^n)\mathrm{K}(\mathbf{x},\mathbf{x}^m,\mathbf{x}^n)}, \label{def_qc} \\
 q_{ijk}^\mathrm{t}&=&\frac{1}{2}\rho\langle c_ic_jc_k\rangle, \label{def_qt} \\
 W^\mathrm{shear}&=&-P_{zx}\frac{\d\langle v_x\rangle}{\d z}, \label{Wshear} \\
 P_{ij}&=&P_{ij}^\mathrm{c}+P_{ij}^\mathrm{t}, \label{def_P} \\
 P_{ij}^\mathrm{c}&=&\frac{1}{2}\overline{\sum_{mn}F_{j}^{mn}(x_i^m-x_i^n)\mathrm{K}(\mathbf{x},\mathbf{x}^m,\mathbf{x}^n)}, \label{def_Pc} \\
 P_{ij}^\mathrm{t}&=&\rho\langle c_ic_j\rangle, \label{def_Pt} \\
  \mathrm{K}(\mathbf{x},\mathbf{x}^m,\mathbf{x}^n)&=&\int\limits_0^1\delta(\mathbf{x}-((\mathbf{x}^m-\mathbf{x}^n)s+\mathbf{x}^n))\d s, \\
 \Gamma^\mathrm{drag}_{ij}&=&-\rho\langle a^{\mathrm{ex}}_ic_j\rangle, \label{Gammadrag} \\
 \Gamma^\mathrm{coll}_{ij}&=&-\frac{1}{2}\overline{\sum_{mn}F_{i}^{mn}(v_{j}^m-v_{j}^n)\delta(\mathbf{x}-\mathbf{x}^m)}, \label{Gamma_coll}
\end{eqnarray}
where the overbar denotes the ensemble average, $\mathbf{F}^{mn}$ the force applied by particle $n$ on particle $m$ when they contact, $\mathbf{x}^n$ the position of particle $n$, $P_{ij}$ the particle stress tensor, $\mathbf{a}^{\mathrm{ex}}$ the particle acceleration due to external forces, and $\mathrm{K}(\mathbf{x},\mathbf{x}^m,\mathbf{x}^n)$ the mathematical expression for a ``delta line'' between $\mathbf{x}^m$ and $\mathbf{x}^n$ with $\delta$ being the delta distribution \cite{Suppl_Mat}. The manner in which we compute these quantities in our numerical simulations is also described in the Supplementary Material \cite{Suppl_Mat}. We note that both $q_{ijk}$ and $P_{ij}$ are separated into two contributions in Eqs.~(\ref{def_q}) and (\ref{def_Pt}), respectively. The contributions with superscript 'c' incorporate the contact forces which the particles experience in collisions and thus encodes the contribution of interparticle collisions, while the superscript 'c' encodes the contributions from the transport of particles between collisions, which is mainly driven by the external forces. We further note that contributions from gravity and buoyancy to $a^{\mathrm{ex}}$ in Eq.~(\ref{Gammadrag}) vanish because $\langle\mathbf{c}\rangle=0$, and thus only the fluid drag contribution to $a^{\mathrm{ex}}$ remains. We finally note that $\Gamma^\mathrm{coll}_{ij}$, additionally to the collisional dissipation of kinetic fluctuation energy into heat, also incorporates the collisional transfers of kinetic fluctuation energy into rotational energy and potential contact energy. However, the latter two contributions are usually much smaller than the former since both the rotational energy \cite{vonPokornyHorender14} and potential contact energy are usually much smaller than the kinetic energy (small interparticle contact times). Moreover, even within the particle bed, where the potential contact energy is significant due to enduring interparticle contacts, it seems likely that the transfer of kinetic energy into potential contact energy can be neglected since the number of contacts of particles approaching each other (contact energy gain) must equal the number of contacts of particles departing from each other (contact energy loss) in the steady state on average. Interestingly, the structures of Eqs.~(\ref{energyx}) and (\ref{energyz}) are the same as the structures one obtains from Boltzmann-like models of granular flows for the particle fluctuation energy balances \cite{Lunetal84,JenkinsRichman85} (i.e., the gradient of the granular heat flux equals the net production of fluctuation energy). However, while such models rely on the assumptions that the particle velocity fluctuation distributions are nearly Gaussian, and that the mean duration of interparticle contacts is much smaller than the mean free time, Eqs.~(\ref{energyx}) and (\ref{energyz}) with their terms given by Eqs.~(\ref{def_q}-\ref{Gamma_coll}) do not. Indeed, these expressions follow strictly from Newton's axioms \cite{Suppl_Mat}. A comparison between Boltzmann-like models of granular flows and an approach like ours can be found in Ref.~\cite{Babic97}.

The sum of Eqs.~(\ref{energyx}) and (\ref{energyz}) describes that a change of granular temperature ($\D T/\D t$) is governed by fluctuation energy production due to granular shear work ($W^\mathrm{shear}$), fluctuation energy dissipation due to fluid drag ($\Gamma^\mathrm{drag}_{ii}$) and interparticle collisions ($\Gamma^\mathrm{coll}_{ii}$), and gradients in the granular heat flux ($\d q_{zii}/\d z$). Considered individually, both Eqs.~(\ref{energyx}) and (\ref{energyz}) have analogous meanings regarding the changes of $T_{xx}$ and $T_{zz}$, respectively. However, it is important to note that, while $\Gamma^\mathrm{coll}_{ii}=\Gamma^\mathrm{coll}_{xx}+\Gamma^\mathrm{coll}_{zz}$ is the collisional granular temperature dissipation rate and thus strictly positive, $\Gamma^\mathrm{coll}_{zz}$ can also be negative due to conversion from $T_{xx}$ into $T_{zz}$ in interparticle collisions.

\section{Simulation results}
\begin{figure}
 \begin{center}
  \includegraphics[width=0.8\columnwidth]{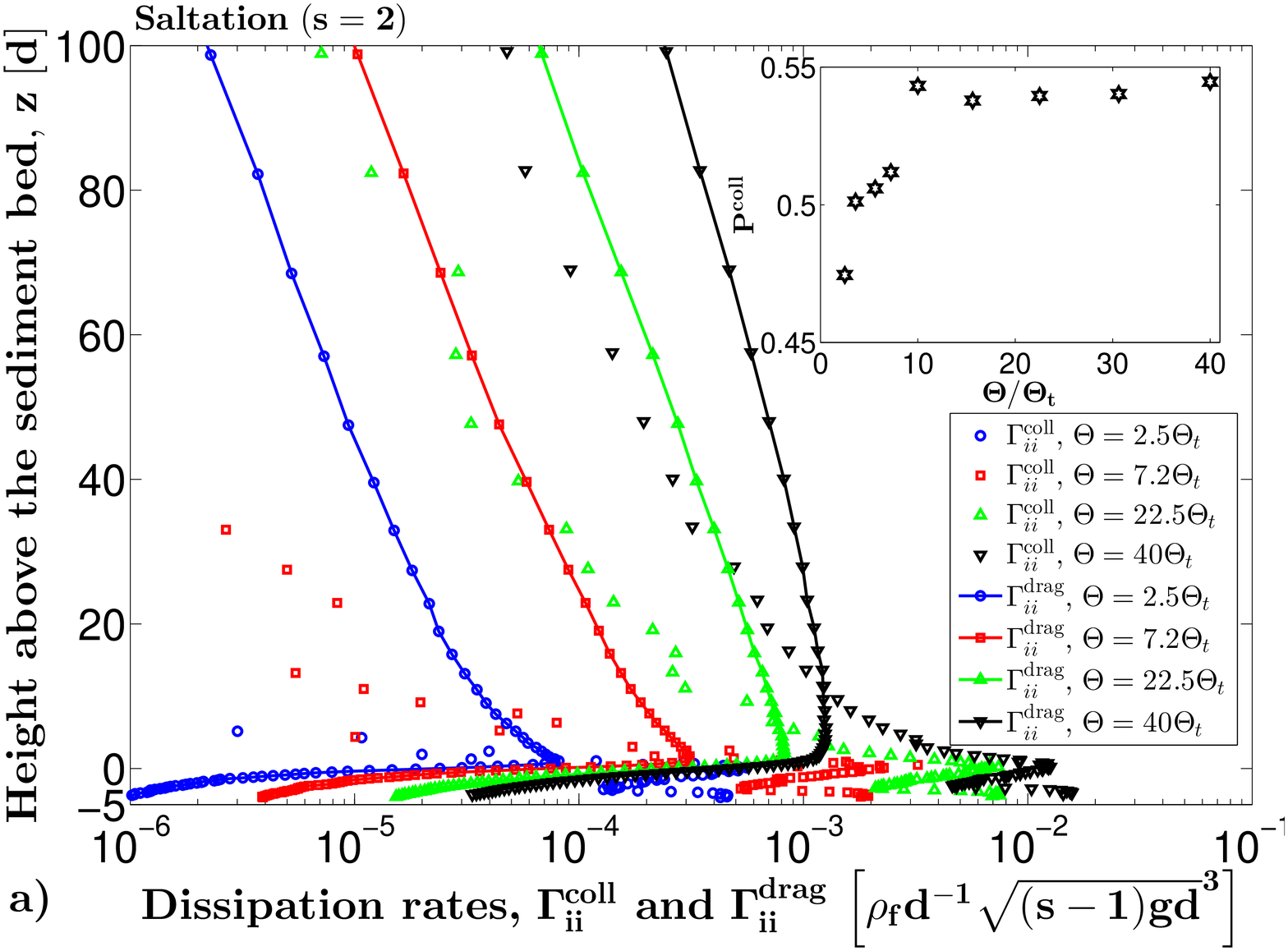} \\
  \includegraphics[width=0.8\columnwidth]{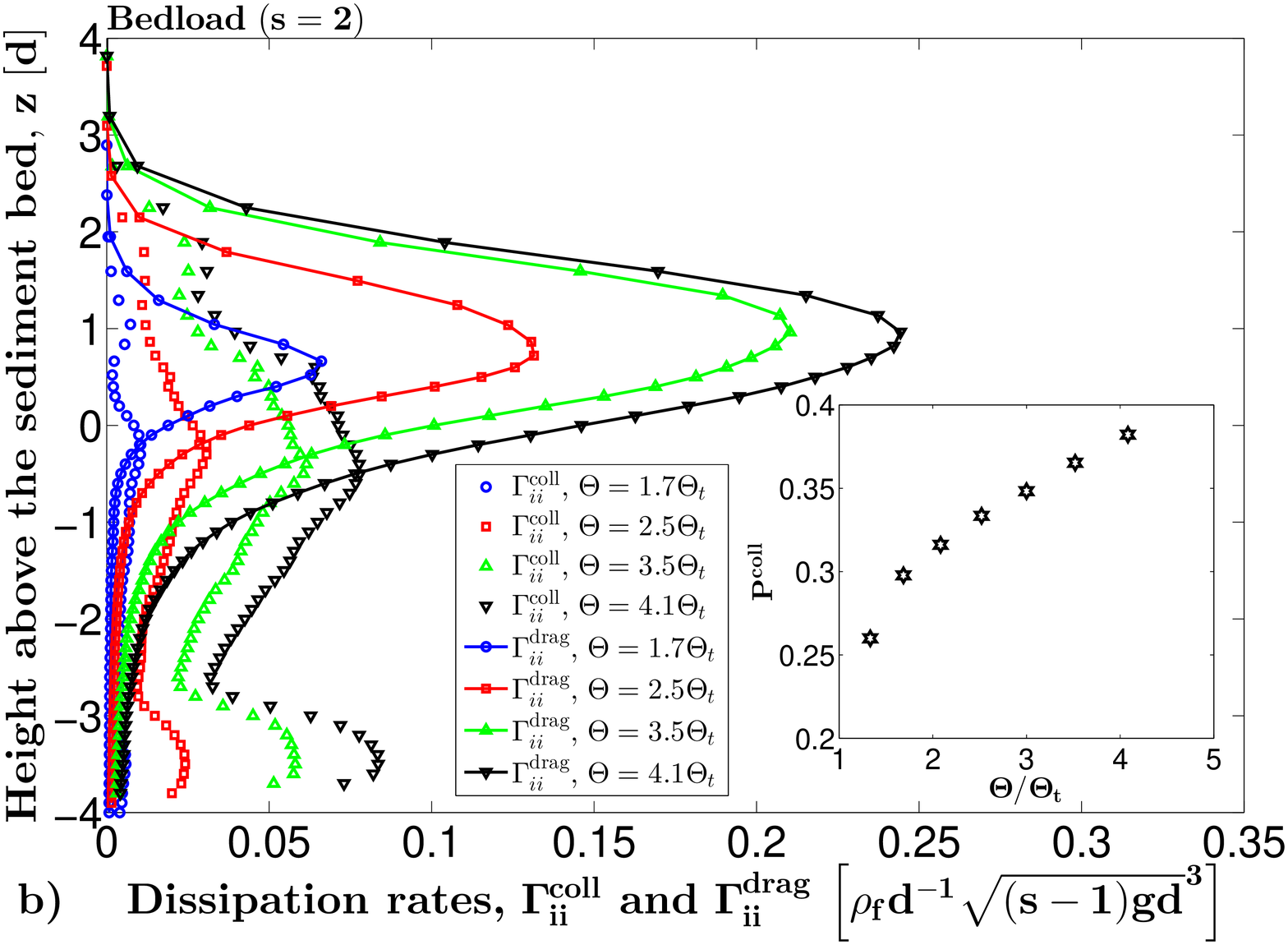}
 \end{center}
 \caption{Vertical profiles of the collisional ($\Gamma^\mathrm{coll}_{ii}$) and drag dissipation rates ($\Gamma^\mathrm{drag}_{ii}$) of fluctuation energy in (a) saltation ($s=2000$) and (b) bedload ($s=2$) for various Shields numbers. The insets show the overall fraction ($P^{\mathrm{coll}}$) of collisional dissipation as a function of the Shields number.}
 \label{Gamma}
\end{figure}

Figs.~\ref{Gamma}a and \ref{Gamma}b display the vertical profiles of $\Gamma^\mathrm{coll}_{ii}$ and $\Gamma^\mathrm{drag}_{ii}$ in saltation and bedload, respectively, for various Shields numbers.
Since dissipation rates are mechanical powers per unit volume (mass density $\times$ acceleration $\times$ velocity), they are plotted in units of $\rho_\mathrm{p}\times\tilde g\times\sqrt{(s-1)gd}=\rho_\mathrm{f}d^{-1}\sqrt{(s-1)gd}^3$, where $\tilde g=(s-1)g/s$ is the buoyancy-reduced value of the gravity, which is the natural unit for particle accelerations. Moreover, in these plots, the height $z=0$ approximately corresponds to the top of the particle bed, defined as the height at which the particle volume fraction reaches one half of its value within the particle bed \cite{Duranetal12}. The region within the particle bed ($z<0$), which has very often been assumed to be immobile in previous numerical studies of fluid-mediated particle transport (e.g. \cite{NinoGarcia98a,KokRenno09}), is of special importance in our study since the effects of particle impacts onto the particle bed are known to extent many layers into the particle bed \cite{Rioualetal03}. In fact, it can be seen in Figs.~\ref{Gamma}a and \ref{Gamma}b that a significant amount of fluctuation energy is dissipated in interparticle collisions within the particle bed. Particularly notable is the region near the plotted simulation bottom ($z\in(-3d,-4d)$), where $\Gamma^\mathrm{coll}_{ii}$ exhibits a local maximum (also in saltation). This maximum is a signature of finite size effects due to external forces applied on the particles from below the plotted simulation bottom.

The most significant result one can extract from Fig.~\ref{Gamma} is that collisional dissipation dominates fluid drag dissipation within and slightly above the particle bed in saltation, while fluid drag dissipation dominates collisional dissipation almost everywhere within the transport layer in bedload. Hence, we computed the overall fractions of collisional dissipation of fluctuation energy in saltation and bedload, defined by
\begin{eqnarray}
 P^{\mathrm{coll}}=\frac{\int\limits_{-\infty}^\infty\Gamma^\mathrm{coll}_{ii}\d z}{\int\limits_{-\infty}^\infty(\Gamma^\mathrm{coll}_{ii}+\Gamma^\mathrm{drag}_{ii})\d z}. \label{Pcoll}
\end{eqnarray}
They are displayed in the insets of Figs.~\ref{Gamma}a and \ref{Gamma}b, respectively, as a function of $\Theta/\Theta_t$. They show that around $45\%-55\%$ of overall dissipation of fluctuation energy is due to interparticle collisions in saltation, but significantly less in bedload.

The results shown in Fig.~\ref{Gamma} are surprising since one would actually expect that interparticle collisions in comparison to fluid drag play a much more important role in bedload than in saltation. One would expect this because the transport layer is much more compressed in bedload, resulting in a much larger interparticle collision frequency and thus collisional acceleration of particles, while the fluid drag acceleration of particles is of the order of $\tilde g$ in both bedload and saltation and thus smaller in bedload than in saltation. The latter follows from a proportionality between average horizontal and vertical forces \cite{Duranetal12}. In fact, even though the drag length ($sd$) is much larger in bedload than in saltation, the average drag acceleration ($\propto (sd)^{-1}V_r^2$) is of the order of $\tilde g$ since the average difference between the flow and particle velocities ($V_r$) is of the order of $\sqrt{s\tilde gd}$ for constant $\mathrm{Re}$.

\begin{figure}
 \begin{center}
  \includegraphics[width=0.8\columnwidth]{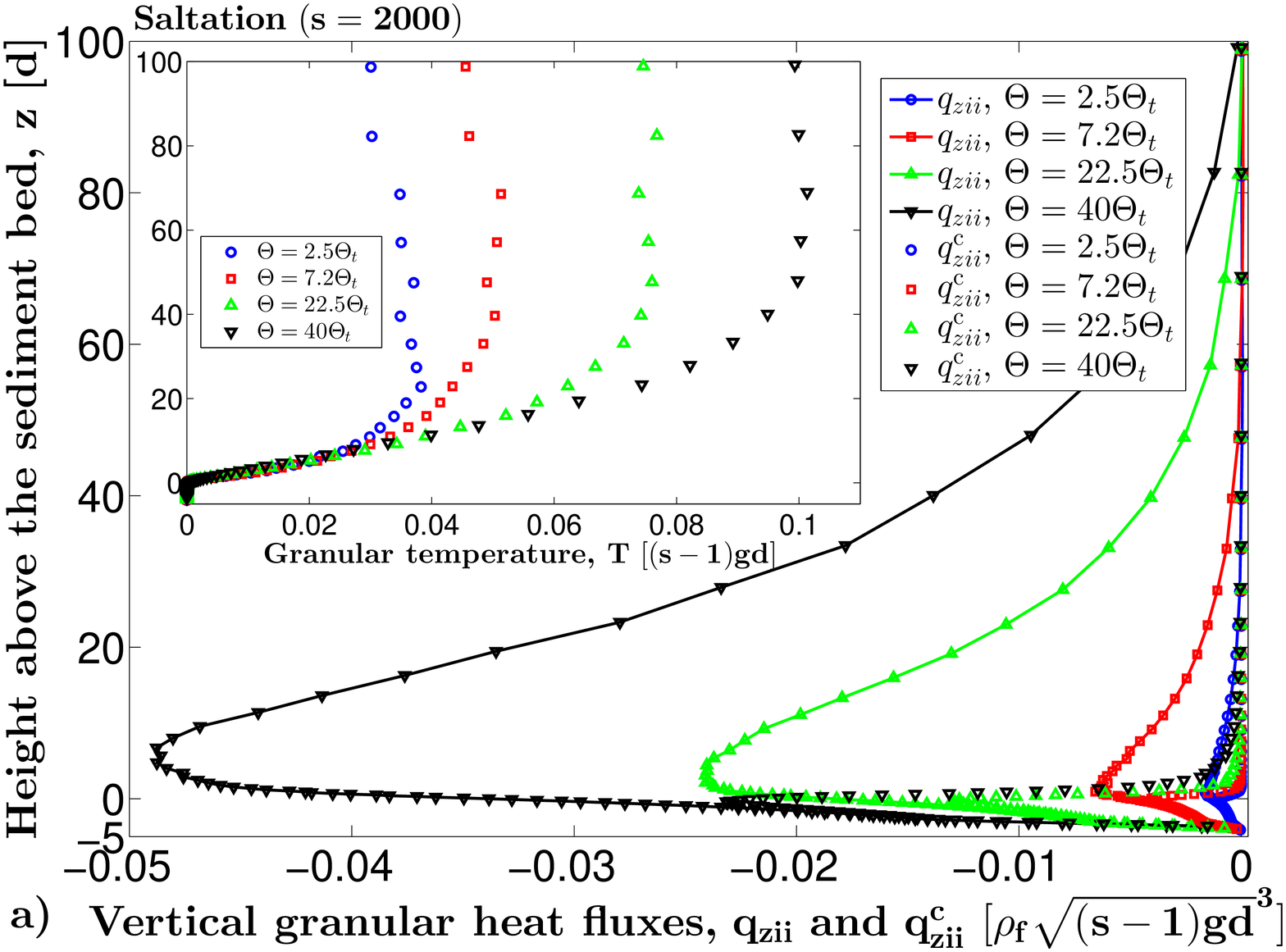} \\
  \includegraphics[width=0.8\columnwidth]{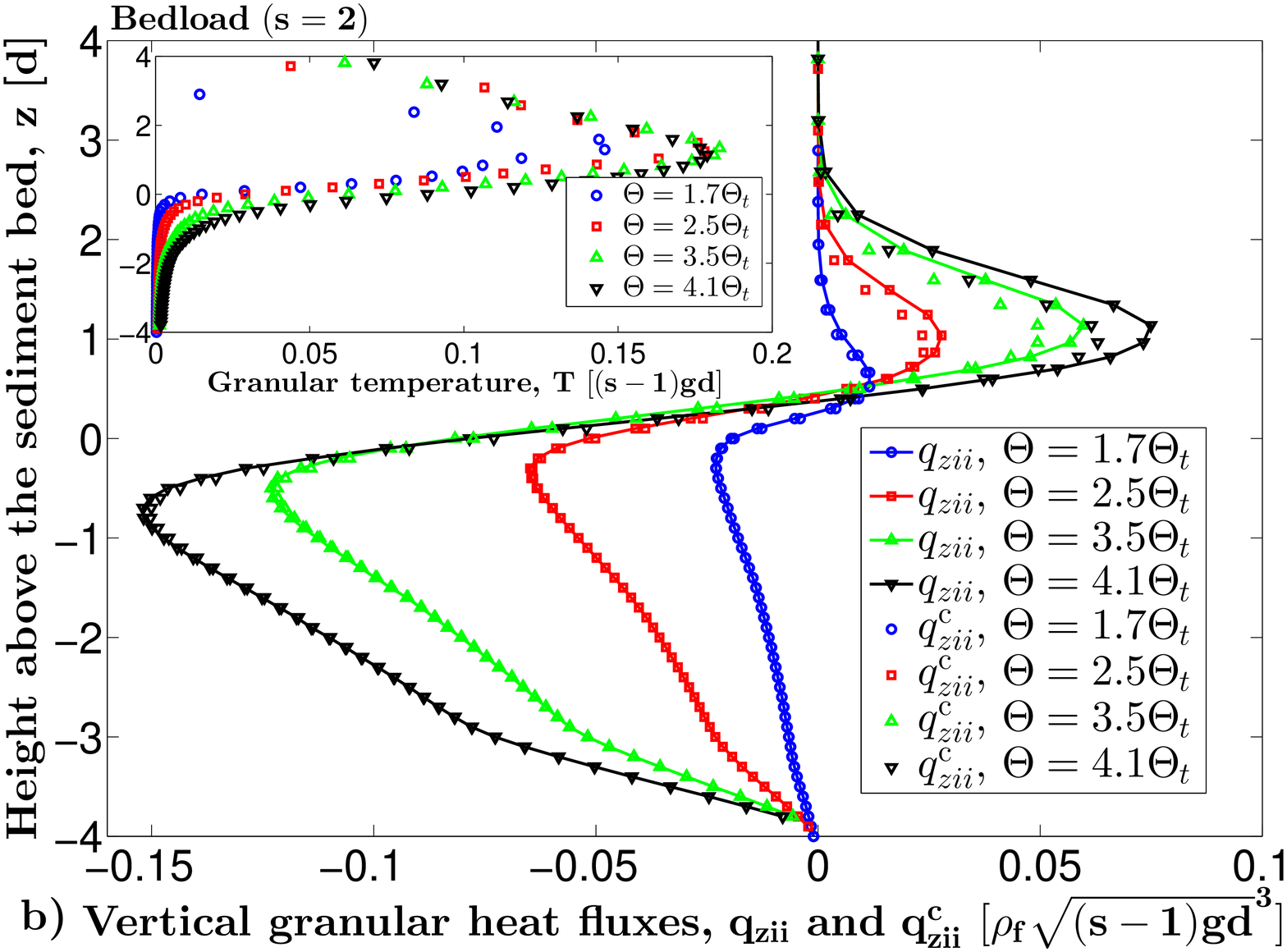}
 \end{center}
 \caption{Vertical profiles of the granular heat flux ($q_{zii}$) and the collisional contribution to the granular heat flux ($q^\mathrm{c}_{zii}$) in (a) saltation ($s=2000$) and (b) bedload ($s=2$) for various Shields numbers. The insets show the vertical profiles of the granular temperature ($T$) for the same Shields numbers.}
 \label{heatflux}
\end{figure}

To further explain why the results shown in Fig.~\ref{Gamma} are surprising, we display in Figs.~\ref{heatflux}a and \ref{heatflux}b for saltation and bedload, respectively, the amount of the granular heat flux ($q_{zii}$) which is due to interparticle collisions ($q^\mathrm{c}_{zii}$) and thus not due to the transport of particles between collisions ($q^\mathrm{t}_{zii}$, see Eqs.~(\ref{def_q}-\ref{def_qt})).
In contrast to the results shown in Fig.~\ref{Gamma}, the results shown in Fig.~\ref{heatflux} are as expected. In fact, due to the compressed (dilute) transport layer in bedload (saltation), the collisional contribution to the heat flux is much larger (smaller) than the transport contribution, which is largely driven by fluid drag. We note that analogous statements can be made for the collisional and transport contributions to the overall particle stress and energy flux tensors (not shown). In other words: While in saltation (bedload) collisions in comparison to fluid drag contribute almost negligibly (dominantly) to the fluxes of momentum, energy, and fluctuation energy, they contribute significantly (almost negligibly) to the dissipation of fluctuation energy.

Fig.~\ref{heatflux} also shows another interesting behavior: Due to Eqs.~(\ref{energyx}) and (\ref{energyz}), positive gradients of the granular heat flux ($q_{zii}$) are associated with a larger production of granular temperature through shear work than dissipation through fluid drag and interparticle collision. Curiously, while $\d q_{zii}/\d z$ is positive for most of the transport layer in saltation (Fig.~\ref{heatflux}a), it becomes negative at $z\approx d$ in bedload (Fig.~\ref{heatflux}b). This sign change corresponds to a strong sign change of the granular temperature gradient occurring at about the same height (inset of Figs.~\ref{heatflux}b), while the granular temperature gradient vanishes for large heights in saltation (inset of Figs.~\ref{heatflux}a). This is just another characteristic of the particle transport layer which strongly distinguishes saltation and bedload.

What causes these fundamental differences in the fluctuation energy dissipation between saltation and bedload? The answer lies hidden in the characteristics of particle impacts onto the particle bed: On the one hand, the order of magnitude of the velocity with which particles in non-suspended particle transport impact onto the particle bed is proportional to $\sqrt{(s-1)gd}=\sqrt{s\tilde{g}d}$ \cite{Duranetal12}. On the other hand, the order of magnitude of the velocity for which particle-bed impacts may induce bed reorganization is proportional to $\sqrt{\tilde gd}$ since the only way in which the ambient fluid significantly influences such effects is via buoyancy. Since the ratio between both velocity scales is equal to $\sqrt{s}$, impacts are the more energetic the larger is $s$. In fact, while particles in bedload ($s=2$) merely rebound from the particle bed, particle-bed impacts in saltation ($s=2000$) are so energetic that they frequently cause a phenomenon known as ``splash'', which describes the ejection of particles from the particle bed \cite{Koketal12}. Consequently, particle-bed impacts in saltation produce much more fluctuation velocity relative to the impact velocity than particle-bed impacts in bedload. This fact can explain the differences in the fluctuation energy dissipation between saltation and bedload, as we will show in the following.

\begin{figure}
 \begin{center}
  \includegraphics[width=0.8\columnwidth]{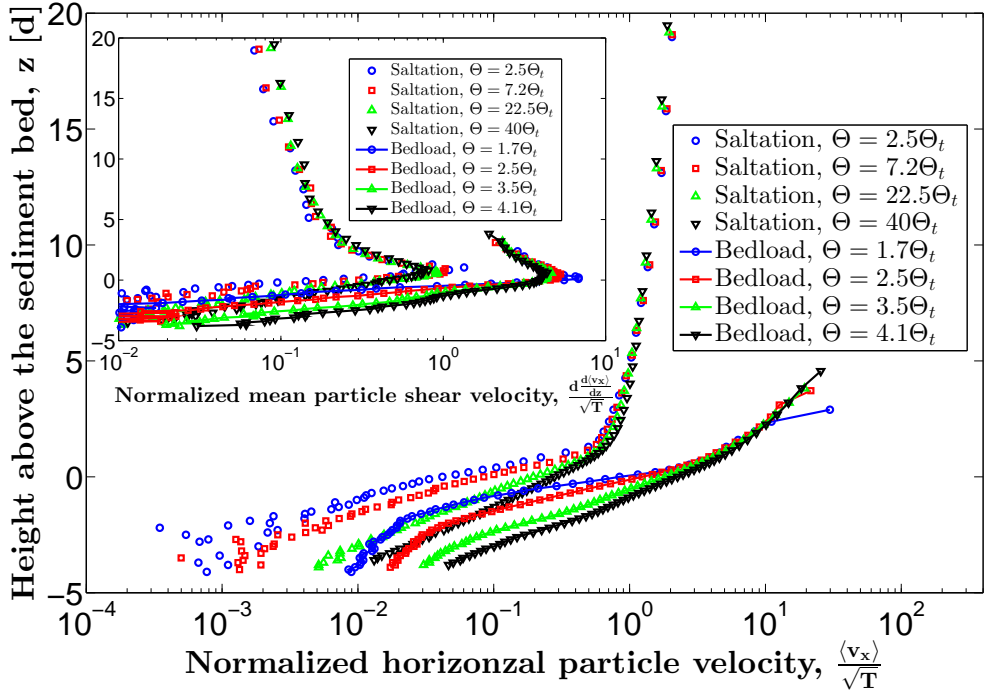}
 \end{center}
 \caption{Vertical profiles of the normalized average horizontal particle velocity ($\langle v_x\rangle/\sqrt{T}$) in saltation ($s=2000$) and bedload ($s=2$) for various Shields numbers. The inset shows the vertical profiles of the normalized particle shear velocity ($\d\langle v_x\rangle/\d z/[\sqrt{T}/d]$) in saltation and bedload for the same Shields numbers.}
 \label{vxdvxT}
\end{figure}

First, as a direct consequence of this fact, it can be expected that the average horizontal particle velocity ($\langle v_x\rangle$) normalized by the fluctuation velocity ($\sqrt{T}$) is much larger in bedload than in saltation in the region near and within the particle bed. Indeed, this is shown in Fig.~\ref{vxdvxT}.
It follows that, in the same region, also the particle shear velocity ($d\d\langle v_x\rangle/\d z$) normalized by the fluctuation velocity ($\sqrt{T}$) is much larger in bedload than in saltation (inset of Fig.~\ref{vxdvxT}). Since the magnitude of particle velocity gradients ($\d\langle v_x\rangle/\d z$) is associated with the magnitude of fluid drag forces, and since the frequency of collisions increases with the fluctuation velocity, the ratio between fluid drag and collisional dissipation should increase with $R=\d\langle v_x\rangle/\d z/[\sqrt{T}/d]$. Hence, since $R$ is much larger in bedload than in saltation due to splash in the region near and within the particle bed, the ratio between fluid drag and collisional dissipation must also be much larger in bedload than in saltation in the same region (see Fig.~\ref{Gamma}).

The strong contribution of collisional dissipation and the significant granular temperature anisotropy in saltation ($T_{xx}/T_{zz}>1$, shown in Fig.~\ref{TxTz}), which exists because granular shear work ($W^{\mathrm{shear}}$) produces horizontal, but not vertical fluctuation energy (see Eqs.~(\ref{energyx}) and (\ref{energyz})), have an interesting side effect:
\begin{figure}
 \begin{center}
  \includegraphics[width=0.8\columnwidth]{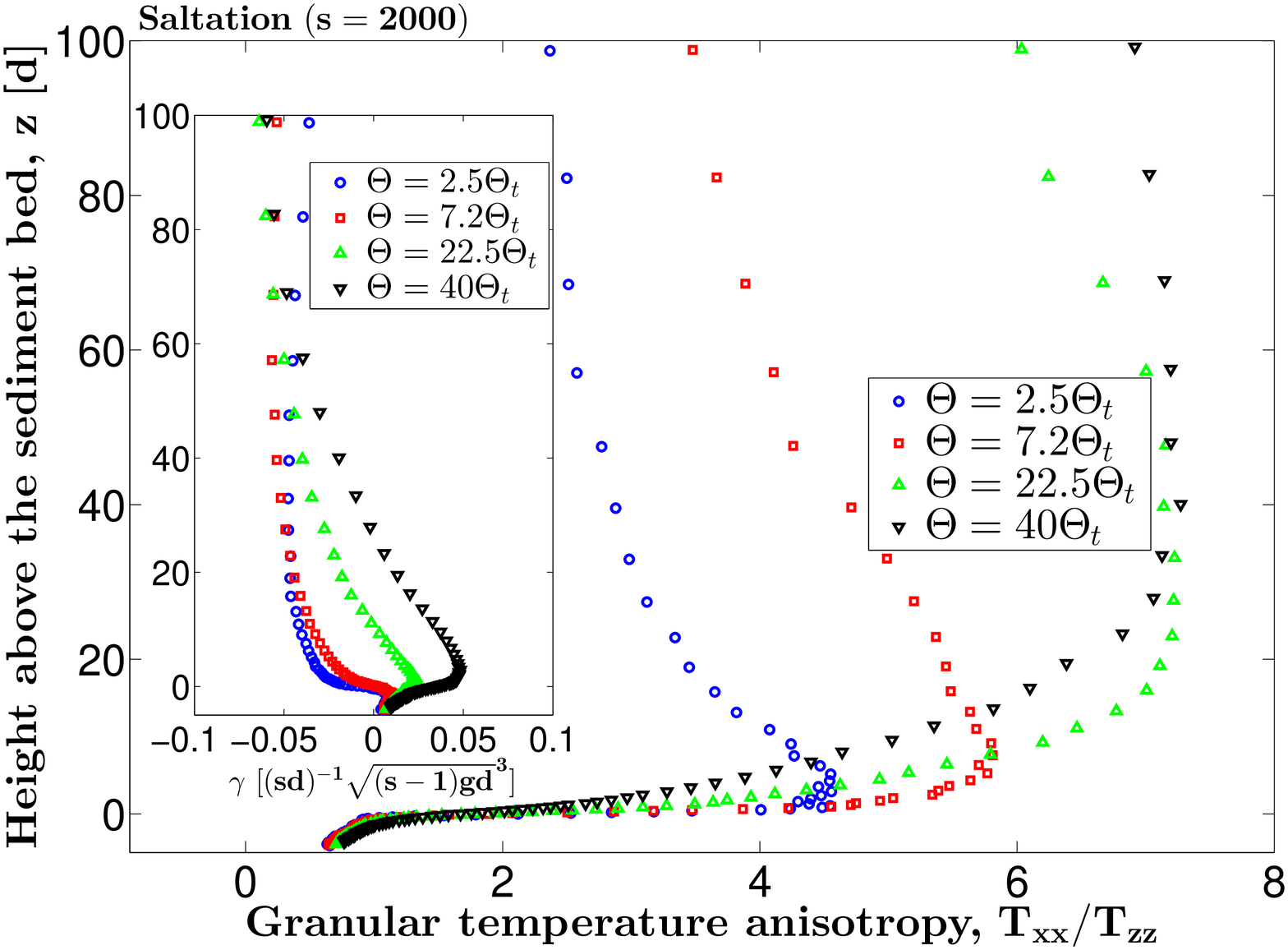}
 \end{center}
 \caption{Vertical profiles of the granular temperature anisotropy ($T_{xx}/T_{zz}$) in saltation ($s=2000$) for various Shields numbers. The inset shows the vertical profiles of $\gamma=-\tilde g\frac{1}{2}\langle v_z^3\rangle/\langle v_z^2\rangle$ in saltation for the same Shields numbers.}
 \label{TxTz}
\end{figure}
Horizontal fluctuation energy is converted into vertical one through interparticle collisions at a rate which is higher than the rate at which vertical fluctuation energy is dissipated by vertical fluid drag forces. This can happen because interparticle collisions tend to make the granular temperature more isotropic, resulting in a production of vertical fluctuation energy in interparticle collisions ($\Gamma^\mathrm{coll}_{zz}<0$), even though the sum of horizontal and vertical fluctuation energy is dissipated through interparticle collisions ($\Gamma^\mathrm{coll}_{ii}>0$). Evidence for this side effect comes from the vertical profiles of $\gamma=-\tilde g\frac{1}{2}\langle v_z^3\rangle/\langle v_z^2\rangle$ (shown in the inset of Fig.~\ref{TxTz}), which exhibit positive values near the particle bed for all Shields numbers. As we explain in the following, in saltation the value of $\gamma$ at a certain height $z$ is approximately equal to $P(>z)=-\int_z^\infty(\Gamma^\mathrm{coll}_{zz}+\Gamma^\mathrm{drag}_{zz})\d z'$, the net production rate of vertical fluctuation energy above $z$, divided by $M(>z)=\int_z^\infty\rho\d z'$, the mass of particles transported above $z$, 
\begin{eqnarray}
 \gamma(z)\approxeq\frac{P(>z)}{M(>z)}. \label{gammadef}
\end{eqnarray}
In other words, $\gamma(z)$ describes the production of vertical fluctuation energy per particle mass averaged over all layers above $z$. Hence, a positive value of $\gamma$ near the particle bed corresponds to a net production of vertical fluctuation energy above the particle bed ($P>0$), which means that the production due to interparticle collisions is larger than the dissipation of vertical fluctuation energy due to vertical fluid drag.

In order to explain this physical meaning of $\gamma$, we use that the major contribution to the granular heat flux and the particle stress tensor at heights above the particle bed comes from the transport of particles between collisions in saltation, as mentioned before (see Fig.~\ref{heatflux}a). Hence, $q_{zzz}\approxeq q_{zzz}^{\mathrm{t}}=\frac{1}{2}\rho\langle v_z^3\rangle$ and $P_{zz}\approxeq P_{zz}^{\mathrm{t}}=\rho\langle v_z^2\rangle$, which follows from Eqs.~(\ref{def_qt}) and (\ref{def_Pt}) and $\langle v_z\rangle=0$. Moreover, we use that the contribution from vertical fluid drag to the acceleration $\mathbf{a}^\mathrm{ex}$ in the vertical momentum balance, $\d P_{zz}/d z=\rho\langle a_z^\mathrm{ex}\rangle$ \cite{Suppl_Mat}, can be neglected (since it is much smaller than the buoyancy-reduced gravity \cite{Duranetal12}), such that
\begin{eqnarray}
 \frac{\d\rho\langle v_z^2\rangle}{\d z}\approxeq-\rho\tilde g. \label{momz}
\end{eqnarray}
Eqs.~(\ref{energyz}) and (\ref{momz}) then allow us to express $\gamma(z)$ as (see Eq.~(\ref{gammadef}))
\begin{eqnarray}
 \gamma(z)=-\tilde g\frac{1}{2}\frac{\langle v_z^3\rangle(z)}{\langle v_z^2\rangle(z)}\approxeq\frac{-\int\limits_z^\infty(\Gamma^\mathrm{coll}_{zz}+\Gamma^\mathrm{drag}_{zz})\d z'}{\int\limits_z^\infty\rho\d z'}=\frac{P(>z)}{M(>z)}.
\end{eqnarray}

\section{Turbulent production of fluctuation energy}
In our simulations, turbulent fluctuations of the flow have been neglected, manifesting itself in fluid drag which strictly dissipates fluctuation energy ($\Gamma^\mathrm{drag}_{xx}>0$, $\Gamma^\mathrm{drag}_{zz}>0$). However, if turbulent fluctuations are taken into account, it is possible that fluid drag might produce instead of dissipate fluctuation energy. This is explained in the following gedankenexperiment:

Let us consider a constant fluid flow (fluid speed $u(x,y,z)=\mathrm{const.}$) without turbulent fluctuations in which each particle moves with the velocity of the flow, corresponding to vanishing fluctuation energy. If we now turn on turbulent fluctuations, the particles will obviously gain fluctuation energy, which means that fluid drag (mean fluid drag dissipation + turbulent production) must net produce fluctuation energy. This production of fluctuation energy will continue until a new steady state is reached (``steady'' hereby refers to $\partial/\partial t=0$ after ensemble averaging). In this steady state, the particles have a certain granular temperature $T_\mathrm{turb}$ at which turbulent production exactly balances dissipation of fluctuation energy due to mean fluid drag (the particle concentration shall so small that interparticle collisions do not occur). Now we externally impose a certain fluctuation energy with $T>T_\mathrm{turb}$ on the particle phase and let the system relax afterwards. In this case, the particles will loose fluctuation energy until the same steady state is reached. This shows that depending on the conditions, the net effect of fluid drag (mean fluid drag dissipation + turbulent production) can, indeed, be both production and dissipation of fluctuation energy.

In order to determine the strength of turbulent production of vertical fluctuation energy in saltation, we compared our DEM simulations with experimental measurements of $\gamma$. In fact, we carried out PTV measurements in a wind tunnel during steady transport of sand for two different sand beds consisting of fine ($d=230\mu$m) and coarse sand ($d=630\mu$m), respectively (see Ho et al. \cite{Hoetal11,Hoetal14} for a detailed description of the experimental setup). From these measurements we obtained
\begin{eqnarray}
 \gamma(13d\pm5d)&=&-\frac{\tilde g}{2}\frac{\overline{\sum\limits_{p:z^p\in(8d,18d)}m^p(v_z^p)^3}}{\overline{\sum\limits_{p:z^p\in(8d,18d)}m^p(v_z^p)^2}}\approx-\frac{\tilde g}{2}\frac{\overline{\sum\limits_{p:z^p\in(8d,18d)}(v_z^p)^3}}{\overline{\sum\limits_{p:z^p\in(8d,18d)}(v_z^p)^2}},
\end{eqnarray}
where the overbar denotes the ensemble average like before, $m^p$ the mass of particle $p$, and the mass-weighted average was approximated by the number-weighted average since it was not possible to determine the masses of the tracked particles with sufficient precision. $\gamma(13d\pm5d)$ obtained from the fine and coarse sand experiments and from the simulations is plotted in Fig.~\ref{expgamma} (the error bars correspond to the $95\%$-confidence intervals) as a function of $\Theta/\Theta_t$, revealing three interesting qualities:
\begin{figure}
 \begin{center}
  \includegraphics[width=0.8\columnwidth]{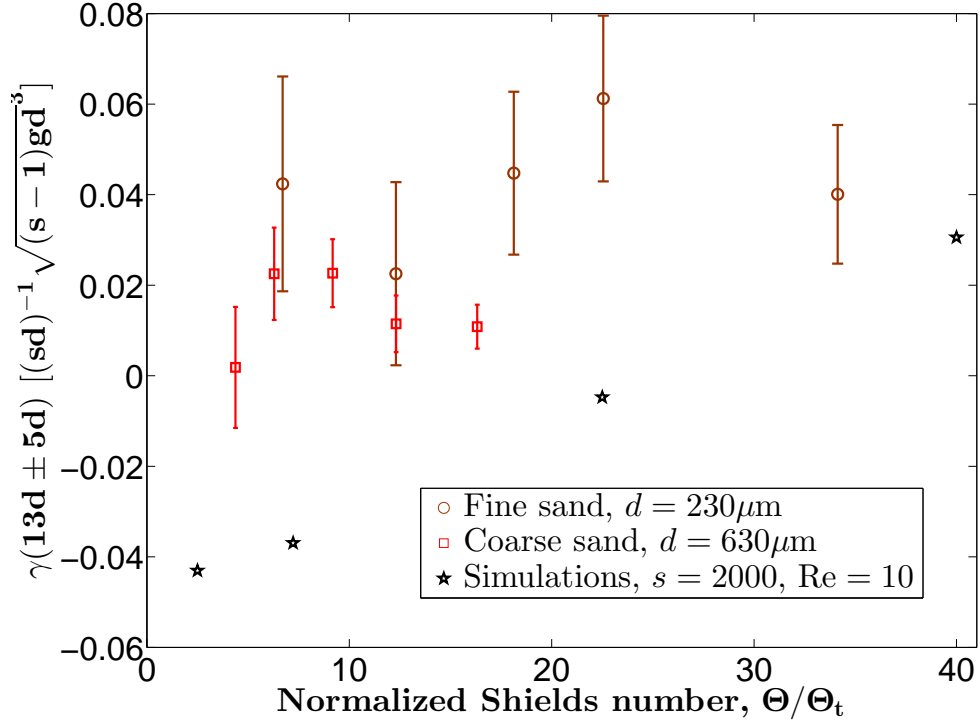}
 \end{center}
 \caption{PTV measurements with $95\%$-confidence intervals of $\gamma=-\tilde g\frac{1}{2}\langle v_z^3\rangle/\langle v_z^2\rangle$ near the sand bed ($z\in(8d,18d)$) in a wind tunnel during steady transport of fine (brown circles) and coarse sand (red rectangles) under a constant air flow for different Shields numbers. The black stars correspond to $\gamma(13d\pm5d)$ obtained from our DEM simulations ($s=2000$, $\mathrm{Re}=10$).}
 \label{expgamma}
\end{figure}
First, $\gamma(13d\pm5d)$ is positive at Shields numbers close to the threshold, where it is negative in our DEM simulations. Second, the values of $\gamma(13d\pm5d)$ are larger in the fine than in the coarse sand experiments, which are in turn larger than in the simulations. And third, the trend in the simulations seems to be stronger than in the experiments. This indicates that turbulent production dominates the vertical fluctuation energy balance at heights larger than $z\approx13d$. In fact, only turbulence can explain why vertical fluctuation energy is produced near the threshold ($\gamma(13d\pm5d)>0$), at which the effects of interparticle collisions are known to be very small at such heights \cite{Koketal12}, since it is the only remaining production mechanism. It is also natural that the effects of turbulence do not change much with the Shields number since the flow shear stress in the saltation layer also does not change much \cite{Koketal12}, which explains the weaker trends with $\Theta/\Theta_t$ in the experiments. Finally, it is also expected that the effects of turbulence are stronger (larger $\gamma(13d\pm5d)$) for fine than for coarse sand \cite{Koketal12}. After all it is the very fine dust which is transported in suspension with the air due to turbulent fluctuations of the flow \cite{Koketal12}.

\section{Conclusion}
The first main finding of our study is that collisional dissipation of fluctuation energy plays a much more important role relative to fluid drag dissipation in saltation than it does in bedload (see Figs.~\ref{Gamma} and \ref{vxdvxT}), even though interparticle collisions transport much less momentum, energy, and fluctuation energy in saltation (see Fig.~\ref{heatflux}). It can be expected that this difference between bedload and saltation is even more pronounced in natural systems because the hindrance effect (i.e., the increase of the mixture viscosity with particle volume fraction \cite{IshiiHibiki11}), which has been neglected in our simulations, results in a significant increase of the drag force with the particle volume fraction ($\phi$). This means the maximum of the fluid drag dissipation rate near the top of the particle bed in bedload (see Fig.~\ref{Gamma}a), where $\phi$ is large, should be even more pronounced, while in saltation with its dilute transport layer, the influence of the hindrance effect is most likely negligibly small. We speculate that another reason why this difference between bedload and saltation should be even more pronounced in natural systems is the lubrication force, which has been neglected in our simulations, as we explain in the following. In fact, from experiments it is known that the coefficient of restitution ($e$) becomes a function of the Stokes number $St=sRev_{\mathrm{imp}}/(9\sqrt{s\tilde gd})$ due to the lubrication force \cite{Gondretetal02}, where $v_{\mathrm{imp}}$ is the relative impact velocity at collision. Only when the Stokes number is sufficiently large ($St>3000$) is $e$ constant. For granular Couette flows under gravity, it is known \cite{ZhangCampbell92} that a decrease of $e$ from $0.9$ to $0.4$ only increased the overall dissipation of fluctuation energy by about $10\%$, although one would expect a much stronger increase ($(1-0.4^2)=4.4(1-0.9^2)$), because of a strong decrease of the granular temperature ($T$) and thus the collision frequency in the dense region near the top of the particle bed. Since granular Couette flows are from all dry granular flows possibly the ones most similar to bedload, we speculate that the main effect of the lubrication force or other factors effectively decreasing $e$ (e.g., increased tangential friction in contacts \cite{JenkinsZhang02}) on bedload is a decrease of $T$. This should affect collisional dissipation more strongly than drag dissipation because $T$ is one of the physical parameters controlling the collision frequency (see our discussion of Fig.~\ref{vxdvxT} regarding the parameter $R$), which explains why the lubrication force should increase fluid drag dissipation relative to collisional dissipation in bedload and thus enhance the different importances of fluid drag and collisional dissipation of fluctuation energy for bedload and saltation in natural systems.

These different importances of fluid drag and collisional dissipation might be of high relevance for attempts to develop a theory which unifies these regimes. For instance, there seems to be a connection between entrainment of bed particles and the manner in which fluctuation energy is dissipated: In bedload, in which particle fluctuation energy is mainly dissipated via fluid drag, entrainment of bed particles mainly occurs due to fluid drag \cite{vanRijn93}, while in saltation, in which particle fluctuation energy is mainly dissipated via interparticle collisions, entrainment of bed particles mainly occurs due to grain-bed collisions \cite{Koketal12}. This connection suggests a further connection between the particle transport and fluctuation energy dissipation rates since the bed particle entrainment rate is closely related to the particle transport rate \cite{vanRijn93,Hoetal14}. It might thus be possible to explain the different scalings of the particle transport rate with the Shields number in saltation and bedload \cite{Duranetal12} in a single unified analytical model through quantifying the fluid drag and collisional fluctuation energy dissipation rates. Such a model would then, for the first time, provide predictions for the particle transport rate in intermediated regimes in which the particle-fluid density ratio ($s$) is neither very small nor very large (e.g., $s\approx100$ \cite{Duranetal12}) and thus significantly improve our understanding of fluid-mediated particle transport.

The second main finding of our study is that turbulent production of vertical fluctuation energy appears to dominate the vertical fluctuation energy balance in saltation, even in saltation of coarse sand. This was suggested by a comparison between experimental measurements and DEM simulations of the average production of vertical fluctuation energy per particle mass above the particle bed (see Fig.~\ref{expgamma}), which is quite surprising since the effects of turbulence in saltation of sand are usually believed to be small, especially for coarse sand \cite{KokRenno09}.

It seems to be a reasonable conclusion that turbulent production also plays an important role in the horizontal fluctuation energy balance and thus in the actual fluctuation energy balance (horizontal + vertical). It might be possible to test this conclusion in future experiments, which would, however, require precise measurements of the vertical profile of the granular shear work ($W^\mathrm{shear}$, see Eq.~(\ref{energyx})).

We acknowledge support from grants Natural Science Foundation of China (NSFC) No. 41350110226 and National Science Foundation (NSF) No. AGS-1358621. We also thank the Laboratoire de Thermocin\'etique in Nantes for the opportunity to perform our wind tunnel experiments.

%

\end{document}